*Article*

# BYOD Security: A Study of Human Dimensions

Kathleen Downer and Maumita Bhattacharya *

School of Computing, Mathematics & Engineering, Charles Sturt University, Albury, 2640, Australia; kathleendowner@gmail.com
* Correspondence: mbhattacharya@csu.edu.au

**Abstract:** The prevalence and maturity of Bring Your Own Device (BYOD) security along with subsequent frameworks and security mechanisms in Australian organisations is a growing phenomenon somewhat similar to other developed nations. During the COVID-19 pandemic, even organisations that were previously reluctant to embrace BYOD have been forced to accept it to facilitate remote work. The aim of this paper is to discover, through a study conducted using a survey questionnaire instrument, how employees practice and perceive the BYOD security mechanisms deployed by Australian businesses which can help guide the development of future BYOD security frameworks. Three research questions are answered by this study: What levels of awareness do Australian businesses have for BYOD security aspects? How are employees currently responding to the security mechanisms applied by their organisations for mobile devices? What are the potential weaknesses in businesses' IT networks that have a direct effect on BYOD security? Overall, the aim of this research is to illuminate the findings of these research objectives so that they can be used as a basis for building new and strengthening existing BYOD security frameworks in order to enhance their effectiveness against an ever-growing list of attacks and threats targeting mobile devices in a virtually driven work force.

## 1. Introduction and Background

Keeping up with the ever-increasing consumerization of smart mobile devices at workplaces, many organizations have embraced the practice of Bring Your Own Device (BYOD), recognizing that it positively enhances business processes. During the COVID-19 era, many businesses which previously rejected BYOD had to embrace it in order to facilitate remote work [1]. Despite the obvious benefits, BYOD, however, inherits and exacerbates risks already associated with traditional computing technology, as well as introducing its own unique risks. These include unauthorised access, data breaches, data loss, or data leak, amongst others. Moreover, sensitive data may continue to remain on ex-employees' devices, opening up avenues for misuse or malicious activities. However, prior literature reviews and research focusing on the current state of BYOD revealed that BYOD security is still an under-researched concept, as it is relatively young compared to other network security issues, which has created a situation wherein companies are resolving security exploits and closing loopholes as they appear [2–4]. Globally spread studies estimated that an average of 75% of organisations permit the use of mobile devices for work. In some sectors, as high as 85% of employees use personal devices to access work-related sensitive information. Still, only approximately 50% of these actively use security measures on mobile devices [5–8]. Around 50% of the businesses enforce any form of BYOD specific security policies [4,5,9–12].

In the context of BYOD security, its human dimension is crucial. Successful BYOD implementation depends on usability from the employee's perspective [13–15]. Furthermore, employees admitted to actively exploiting loopholes when they disagreed with BYOD security policies or had trouble in using applied security methods [16–18]. The importance of end users must be prioritised, as they determine and maintain the most control over the success of BYOD security frameworks. End users can defy BYOD security initiatives by simply ignoring or being completely unaware of them, exploiting loopholes, or refusing to participate due to diverging opinions. Our research draws attention to this aspect, as businesses should be aware of these implications in order to strengthen countermeasures towards threats targeting BYOD initiatives. Approaching security from the end user's perspective assists businesses by revealing vulnerabilities which can help in the detection of potential internal threats through providing knowledge of employees' device usage habits and feelings towards applied security methods [12,19–21]. This can strengthen BYOD security by alerting businesses about specific threats which require countermeasures to prevent future security breaches and provide more thorough guidelines for end user agreements and usage agreement policies.

This literature review revealed that, despite its importance, there is still a dearth of research that focuses intensively on the human dimensions of BYOD security, especially the end users' perspective, where all participants have actual BYOD experience. This study aims to address this research gap. In this study, we investigate how BYOD security is handled by Australian businesses with a specific focus on end users. This study differs from previous comparable studies regarding BYOD security as it places particular emphasis towards the use of BYOD security frameworks and solutions in order to find out how security methods are actually interpreted and practiced by Australian employees in real life business operations. The specific research questions (RQs) of our current study are as follows:

*RQ.1* Are Australian businesses and employees sufficiently aware of BYOD security aspects in terms of risks involved (threats and attacks) and security mechanisms available to protect mobile devices?

*RQ.2* How have employees responded to security mechanisms enforced by organisations?

*RQ.3* Are there any potential weaknesses in Australian business internal networks from the perspective of end users which have a direct effect on BYOD security and what are they?

To address the above research questions, a mixed approach study conducted through an online survey targeting users of organizational BYOD security initiatives presents valuable information for the following research aims (RAs):

*RA.1* Understand the security practices utilised by Australian businesses to protect resources and staff engaging in BYOD.

*RA.2* Investigate how users practice BYOD security in regard to work, including typical working conditions and how work tasks are completed with mobile devices.

*RA.3* Explore users' perceptions, opinions, and reactions towards organizational BYOD security measures.

The results of this study are expected to inform recommendations for improving future BYOD security frameworks which will enhance usability for intended users and may inspire new approaches in order to improve the protection of mobile devices involved in future corporate BYOD initiatives.

The rest of this paper is organised as follows: Section 2 presents a review of relevant literature; Section 3 outlines the research method used; Section 4 details the results obtained in this study and Section 5 presents some analysis and discussion of the results; Section 6 sheds light on the limitations of this study and finally some concluding remarks are presented in Section 7.

## 2. Literature Review

Incorporating BYOD into an organisation offers ample advantages including enhanced productivity, efficiency, reduced infrastructure expenses, and increased employee morale [7,22,23]. Approximately 70% of businesses are already utilizing some aspects of BYOD and agree that they have experienced improvements in everyday work processes [5,22,24–26]. However, ever-increasing threats and attacks aimed towards portable devices, coupled with the low adoption of BYOD-specific security strategies by organisations and noncompliance by end users leave the BYOD landscape vulnerable.

The majority of the existing literature on BYOD security primarily deals with threat analysis and mitigation strategies without an intensive focus on the human dimension. Palanisamy et al., in [7], presented a comprehensive review of literature on compliance to BYOD policy. The authors concluded that 'threat severity' and 'self-efficacy' were the strongest predictors of BYOD security policy compliance intentions and behaviour. This review also observed that there is a lack of emphasis on security policy in existing studies.

Bello et al. [27] used a case study approach wherein data were collected using a survey questionnaire and interview instruments to determine BYOD usage patterns and perceptions, BYOD security and privacy, and support for BYOD, etc. Some of the limitations of this study include its small sample size and the potential misrepresentation of information by participants in interviews. Tu et al. [28] presented a theoretical model to identify the key factors that influence an employee's intention to comply with organisations' BYOD security policies. The study also observed that the use of a myriad of different types of devices makes it difficult to implement BYOD security effectively. Palanisamy et al. examined users' attitude on the perceived security of enterprise systems (ES) mobility using a questionnaire survey undertaken in China [29]. Their study attempted to explore users' attitude towards different types of security issues, such as mobile device security, wireless network security, computing security, and so on. This study is not specifically targeted at BOYD security, although 93% of the participants' mobile devices were owned by users and are thus relevant in the current context. The data sample used in this study may not be an accurate representative of the wide range of mobile enterprise systems (ES) users in China.

Wani et al. [30] suggested mitigation strategies relevant to BYOD security challenges based on existing security frameworks. Mitigation strategies proposed in this study are limited by two existing security frameworks and were not informed by actual studies undertaken by the researchers. Yang et al. in [31] investigated the factors that affect employees' opt-in decisions with BYOD security policy using an experimental survey. The study concluded that positive BYOD security policy justification framing and post-task security policy exposure are likely to positively influence opt-in decisions and compliance to BYOD security policies by employees. The data collection for this study was restricted to a specific city in China. Moreover, the results of the study have some limitations, as over 61% of the participants did not have a BYOD policy in their organization and communicated their views related to opt-in decisions without real experience. Another study [32] observed, besides positive factors, that management should also take into account environmental factors that may influence employees' security behaviour. According to yet another recent study [33], the factor 'threat severity' seems to have an insignificant impact on BYOD policy compliance behaviour among employees.

Chigada and Daniels in [34] investigated the security implications of BYOD in financial organisations using a qualitative study. This study revealed the prevalent absence of BYOD policies. The study participants were purposefully chosen from the information technology (IT) and IT security departments of the organisation and did not examine general end users' perceptions. Aguboshim and Udobi [35] presented a review of literature on BYOD security issues. Downer and Bhattacharya [36] also presented a comprehensive review of BYOD security challenges. Agudelo-Serna et al., in [37], suggested that an integration of social, technical, organisational, and also environmental factors need to be considered while addressing security issues due to the mobile device usage patterns.

Cho et al. [38] explored the factors affecting employees' adoption intention of BYOD using a survey instrument. Similar to other comparable studies [7], this research concluded that BYOD adoption is primarily influenced by threat and coping appraisal, while other positive inducements, such as organisational commitment and job security, are also important. As the study focused on BYOD adoption intentions, the target population included employees who were thinking of adopting BYOD for their work and, hence, did not capture the perceptions of employees who already had the BYOD experience. Crossler and Bélanger, in [39], observed that the perception of ownership is likely to motivate people to be more protective of their mobile devices. However, this may also make it more difficult for organisations to implement their BYOD security policies. This study focused on measuring actual behaviour instead of intended behaviour. Gbouri and Mensch [14] presented a research plan to examine users' acceptance in adopting a BYOD strategy through a qualitative study using the Unified Theory of Acceptance and Use of Technology (UTAUT) as the basis for aligning interview questions. To our knowledge, the findings of this study are yet to be published at the time of writing this paper.

Chen et al. [40] examined whether and how employees decide to adopt BYOD practices when faced with information security-related conflicts. This study suggested that information security-related conflicts cause information security fatigue among employees. With an increase in such fatigues, employees are less likely to adopt appropriate BYOD practices. Information security fatigue may also negatively influence BYOD adoption by employees. This study was focused on the impact of information security-related conflicts on employees' BOYD decisions, and other factors that influence such decisions were not considered. Michelberger and Fehér-Polgár [12] identified and classified risks to employees and employers associated with BYOD in order to inform corporate BYOD policy and procedure frameworks. This study relied on existing literature for the relevant information.

Palanisamy et al. [16] reviewed seventeen relevant articles and presented an overview of risks that arise from employees' security policy noncompliance behaviour. In another review article [2], Ratchford et al. presented a review of thirty-eight articles, specifically to identify BYOD security issues. The majority of their reviewed articles focused mainly on IT domain-related issues. Their findings also show that the most frequently addressed concerns are BYOD security issues corresponding to policies. Weidman and Grossklags [13] investigated employee acceptance or opposition to a mandated new authentication system utilizing employee-owned mobile devices. Although this study is not relevant in the context of BYOD security, the findings may shed some light on employees' BYOD adoption intentions.

In Table 1, we also present the main focus and limitations of some of the influential older (pre-2016) literature relevant to BYOD security. As is apparent from this literature review, studies that intensively focus on BYOD-experienced end users' perceptions, with a focus on security frameworks, are rare. This research gap is addressed in our study.

Table 1. Focus and limitations of some of the key older (pre-2016) publications.

| Research | Focus | Limitations | Review (R)/Investigation (I) |
| --- | --- | --- | --- |
| French et al., 2015 [22] | Comparison of BYOD usage worldwide; mention of BYOD strengths and security challenges. | Superficial discussion of end users; no proposed recommendations. | R |
| Garba et al., 2015 [25] | Case study comparing BYOD security between technological, educational and financial organisations; introduction of a few attacks and threats targeting mobile devices and a summary of useful BYOD policies. | No discussion of existing security mechanisms for BYOD, despite discussions about how organisations handle security and policy suggestions. | R |
| Bradford Networks, 2012 [41] | Explanation of security challenges and general guidelines for forming BYOD policies. | Limited explanation about how to enhance access control solutions. | I |
| Eslahi et al., 2013 [42] | In depth discussion about MDM, MIM, and MAM, and Mobile bot nets. | Limitations of MDM, MIM, and MAM are not mentioned. | R |
| Gajar et al., 2013 [43] | Considerations and background information useful for understanding BYOD and some information about MDM and common security challenges. | The focus revolves exclusively around access control techniques for securing BYOD devices. | R |
| Leavitt, 2013 [44] | Explains mobile-specific security frameworks, cloud storage, and a few mobile device vulnerabilities. | Only acknowledges a few threats and MDM-related end point security methods. | I |
| Morrow, 2012 [4] | Mobile device vulnerabilities as challenges, supported by statistics that emphasise the weight of these issues. | Information about security is heavily influenced by statistics, thus is biased by trends reported years ago. | I |
| Romer, 2014 [3] | Explanation of access control methods that protect data from some threats and attack types. | Relatively limited number of security threats and attacks are presented, and advice given revolves around access control initiatives. | I |
| Scarfo, 2012 [45] | Presentation of trends and security frameworks favoured by businesses. | Biased towards desktop virtualisation, in comparison to other solutions presented. | I |
| Tokoyoshi, 2012 [46] | Explores issues that influence BYOD policies and offers suggestions to mitigate the risks. | Security frameworks are mentioned yet are not explained in detail. | I |
| Disterer et al., 2013 [47] | Opportunities and risks of BYOD and comparison of desktop virtualisation models. | Only discusses desktop virtualisation models with a mere mention of MDM. | I |
| Wang et al., 2014 [26] | Specific security frameworks and a wide range of attack types and challenges are discussed. | Frameworks suggested are limited to VPNs and MDM-based variations. | R |

## 3. Methods

The aim of the research is to gain an informed vision of user perceptions and practices of BYOD security in Australian business which would reveal some trends and usage patterns affecting resource and data protection and privacy. Background knowledge of employees' working habits can help organisations modify existing or design new BYOD security frameworks to counteract potential threats and vulnerabilities. The study uncovers details about the relationship between security methods and the user interactions with them, which clarifies why users demonstrate certain behaviours towards specific security measures. This exposure could discourage and aid the prevention of internal security breaches, as well as spawn ideas to enhance future BYOD frameworks. An awareness of feelings and actions of employees can help businesses formulate security frameworks that provide a better compromise between business needs and employees, enabling the deployment of security measures with higher employee acceptance.

The study was conducted through an online survey, available to end users who fit the criteria of working through their mobile devices as part of their employment conditions. The survey was anonymous, and the questions were formulated to collect information concerning employee experiences with corporate BYOD security initiatives. Survey respondents varied widely in terms of experiences with BYOD and in their personal knowledge of mobile device security, which consequently affects the protection of company resources, returning interesting results for this research.

The survey was hosted on the website 'Survey Monkey' and was distributed via the following methods: publicly shared on social media websites such as Twitter and Face-book, online professional networking websites, and several organisations were also directly approached through researchers' personal professional network. The survey was left open for three months to facilitate data collection.

*3.1. Survey Questionnaire Design*

The survey questions asked were divided into four distinct sections. An overview of each section's focus, aims of asking certain questions, and how the provided responses are expected to answer research objectives, are explained below.

**Section 1. Fundamental demographic information**. This section collects fundamental statistics such as preferred devices and determines how reliant modern organisations are towards BYOD strategies. Basic information recorded includes descriptions of working conditions, career paths, business size, reliance of employees on BYOD initiatives, and favoured hardware and operating systems. These factors influence how BYOD is handled by companies and potentially uncovers which career paths tend to be more reliant on these initiatives, as well as why and how.

**Section 2. Business BYOD security practices.** This section focuses on the security practices applied by the organisation that respondents belong to, how security measures are enforced, and the technologies and governance policies that are utilised and how they relate to everyday operations.

**Section 3. User practices in regard to BYOD at work.** This section is concerned with how employees engage in BYOD initiatives on a daily basis. The aim is to find out which work tasks are typically completed using mobile devices and how employees understand, accept, and adhere to security measures affecting BYOD practices.

**Section 4. User perceptions of BYOD in the workplace.** This section is concerned with people's personal experiences, feelings, reactions, and opinions on how organisations manage BYOD and enforce associated security practices in the workplace. The final open-ended question in this section reveals insightful information for this study, as it gave respondents the opportunity to offer feedback on how BYOD security could be improved in their own workplace.

*3.2. Data Analysis*

The quantitative component of the data has been analysed using statistical software, (IBM SPSS® Statistics campus edition) whilst thematic analysis was used to analyse the qualitative component of the data. Survey questions formatted as multiple choices were measured quantitatively, while open-ended questions which yielded unique responses are considered qualitative. Pivotal quantitative collections of data are presented graphically throughout Section 4, Survey Results. Examining the relationship between user experiences against the security measures used by organisations reveals previously overlooked or unknown weaknesses of BYOD security frameworks and clarifies the importance of end user interactions and opinions.

This study was approved by Charles Sturt University's Human Research Ethics Committee.

**4. Survey Results**

In this section, data collected through the survey is presented to expose patterns and trends in how users interpret policies, interact with BYOD security mechanisms, and how their behaviour affects those methods applied by organisations to protect resource and data assets that are involved in BYOD initiatives.

*4.1. Fundamental Demographic Information*

Fundamental statistics collected in the first section of the survey describes environmental factors that influence how employees react to workplace BYOD initiatives. Company culture, size, parent industry, and mobile hardware choices equally affect the success of BYOD in the workplace.

This study targeted 250 participants who met the criteria of working through their mobile devices as part of their employment conditions in Australian organisations. Respondents came from a variety of industries, ranging from telecommunications, finance, government sector, law, education, engineering, hospitality, and retail, from both public and private sectors in the Australian workforce. The recruitment strategy ensured that bias towards any specific industry was unlikely. The organisations referenced throughout the survey varied in size, global distribution (54% maintain offices in multiple countries), and dependence on information technology to drive processes, as summarised in Table 2.

**Table 2.** Comparison of environmental factors and hardware influencing BYOD security.

| Factor | | | | | |
|---|---|---|---|---|---|
| Number of Employees | <100 | 100–250 | 250–500 | 500–1000 | >1000 |
| | 42.86% | 11.43% | 11.43% | 2.86% | 31.43% |
| BYOD Timeline | <1 year | 2–3 years | 3–5 years | >5 years | Unsure |
| | 11.43% | 5.71% | 17.14% | 37.14% | 28.57% |
| Working Conditions | Permanently remote | Daily travel to clients/sites | Sometimes out of office | On call after hours | Permanently in office |
| | 2.86% | 5.71% | 31.43% | 20% | 40% |
| Mobile Devices | Smartphone | Tablet | Laptop | Desktop PC | Other |
| | 91.43% | 11.43% | 54.29% | 20% | 2.86% |
| Operating Systems | Android | Apple iOS | Windows 8 | Windows 10 | Other |
| | 54.29% | 34.29% | 14.29% | 28.57% | 2.86% |

The majority of Australian businesses referenced in this study have been engaging in BYOD for over 3 years, and it is expected that this will continue to rise. Currently, smartphones are the largest driving force behind BYOD initiatives in Australian workplaces, followed by laptops, as they are easy to use, affordable, lightweight, and are used for both telecommunications as well as computing abilities. It is important to note that some people may use more than one mobile device when participating in BYOD. As Android is the preferred operating system for modern businesses, it is logical that more malware is aimed towards it than its competitors because of its popularity [42,44,48]. Android and Apple iOS have and likely will continue to dominate the market [49,50], thus businesses would benefit from catering security mechanisms to prioritise their unique needs [43].

As nearly 50% of Australian employees use their privately owned mobile devices for work (See Table 3), this is potentially where a lot of security issues reside, which is consistent with publications claiming that 50% of employees will extend the usage of their privately owned devices for work [5,6]. Clearly, a majority of employees prefer to use their own devices regardless of their physical location (remote or in office) [5,22,23,51]. Allowing employees to use personally owned devices improves productivity and promotes a pleasant work environment [23,52,53]. Reliance on employees' willingness to provide hardware for work impacts how evasive business applied security policies can be, according to local privacy laws.

**Table 3.** Reliance on mobile devices engaged in BYOD initiatives.

| Factor | | | | |
|---|---|---|---|---|
| Use of privately owned devices | Always | Sometimes | Rarely | Never (company-provided device) |
| | 48.57% | 34.29% | 5.71% | 11.43% |
| Dependence on devices for work | Extremely | Highly | Moderately | Casual use |
| | 25.71% | 28.57% | 34.29% | 11.43% |

Two key external factors influence how BYOD is applied in professional environments, the security methods deployed, and who participates in BYOD initiatives, which are: employee's job roles and the industries that the business primarily defines itself by (results are predisposed by employee's perceptions).

In Australian businesses, BYOD is favoured more by managers, followed by IT professionals. Possible reasons for this may be that additional lines of communication enabled by mobile devices give managers greater control over monitoring teams, delegating tasks, and keeping updated on project progress. IT professionals' reliance on BYOD is perhaps due to the virtually uninterrupted connectivity afforded by BYOD, which significantly assists them in carrying out their job. In these occupations, where high connectivity is important, employees have occasionally relied on BYOD even when multiple devices were given by their employer. Retail and telecommunications industries are pioneering the direction of BYOD in Australian workplaces, which may be a side effect of these industries being key providers of the hardware (e.g., smartphones) and services (data, phone plans, etc.) on which BYOD is dependent. Government organisations are also starting to gain momentum, as the Australian government has actively enforced and set examples for subsequent laws such as the Privacy Act 1988 for BYOD security [54]. Table 4 presents some statistics regarding relationships among occupation, industry, and BYOD usage as observed in our study.

**Table 4.** Organisations' parent industries and respondents' job role classifications.

| Job Roles by Classification | | | |
|---|---|---|---|
| Management | Information Technology | Retail | MISC |
| 42.85% | 34.31% | 11.43% | 25.74% |
| Industries | | | |
| Retail | Telecommunications | Education | Government Sector |
| 22.86% | 14.29% | 5.71% | 11.43% |
| Health Care | Finance | Media and Arts | Information Technology |
| 8.57% | 5.71% | 11.43% | 8.57% |
| Law | Hospitality | Engineering | Not for Profit |
| 2.87% | 2.87% | 2.87% | 2.87% |

*4.2. Business BYOD Security Practices*

Discovering which security mechanisms are currently used for BYOD presented promising results, which suggests that Australian businesses closely follow the security standards applied and spread globally by other businesses.

Figure 1 shows that the most popular security mechanisms are those that already exist and were trusted methods in business networks prior to the introduction of mobile devices (especially smartphones), made evident by the popularity of antimalware, network access controls, and desktop virtualisation models.

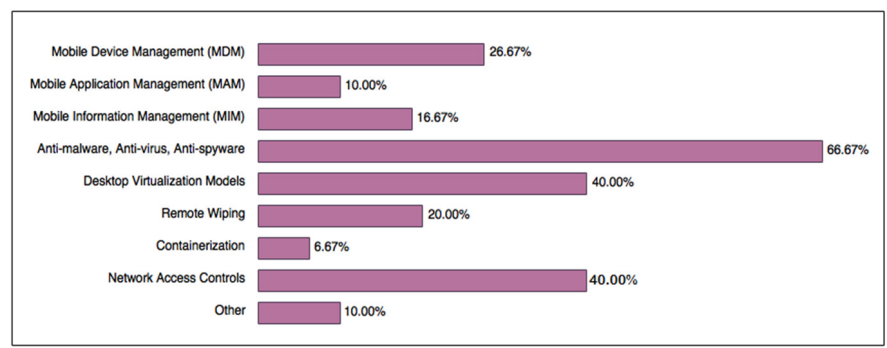

**Figure 1.** Active BYOD security mechanisms in Australian organisations.

Mobile Device Management (MDM) is a multi-functional framework which grants businesses the ability to strictly control mobile devices. MDM solutions contain a main component which manages protocols, provides constant control and monitoring, resides within the company's network, and relies on the exchange of certificates to authenticate and communicate with MDM agents, which are installed on mobile devices [36]. While there are signs that MDM and its variations are becoming more popular for protecting devices regularly involved in BYOD activities, there is still a resistance to, or disregard for, alternative methods such as remote wiping and containerisation. These low usage rates (20% and 6.67%, respectively) suggest that businesses are not as concerned with, or are naive towards, their power in avoiding cases of data leakage, contamination [55], and internal exploitation, as they are in preventing outsider access to company data.

Mobile Application Management (MAM) is a flexible alternative to MDM, as the scope of protection concerns a specific set of applications on the mobile device. MAM allows the company to apply security policies, lock down, define access control rules, configure software behaviours, remote wipe applications under its control, restrict access to unauthorised applications, and install approved applications [36]. On the other hand, Mobile Information Management (MIM) is primarily concerned with data integrity and encryption, determines application and personnel access, and ensures document synchronization amongst multiple devices, whilst simultaneously administering security procedures such as malware scanning [36]. Among businesses using mobile management mechanisms, an equal number tend to choose either MDM or MIM and MAM.

Other security mechanisms mentioned by survey respondents included network firewalls, explicit rules in usage agreements concerning authorisation, and policies that check if the device contains certain encryption rules before allowing data exchanges between it and the business network.

Approximately 40% of Australian businesses included in this study have established formal BYOD policies in place, such as signed acceptable usage, user agreement policies, and liability contracts. Of these, 6.9% update their policies yearly, 20.26% update contracts every few years, and 13.79% only update contracts when major changes occur. One of the most alarming statistics was that the remaining 60% of end users claim that there are no formal agreements addressing BYOD security, despite the strong presence of personal mobile devices in the workforce. Companies who provide training and support for protecting devices engaged in BYOD, along with emphasising the importance of reducing threats and attacks targeting mobile devices, are aligned according to survey results and are highlighted in Figure 2. Organisations need to educate employees and extend technical support as one of the main aims for BYOD security frameworks. Almost 50% of organisations involved in this study do not currently provide any form of training about best practices for BYOD security, which could be linked to a lack of formal BYOD security policies.

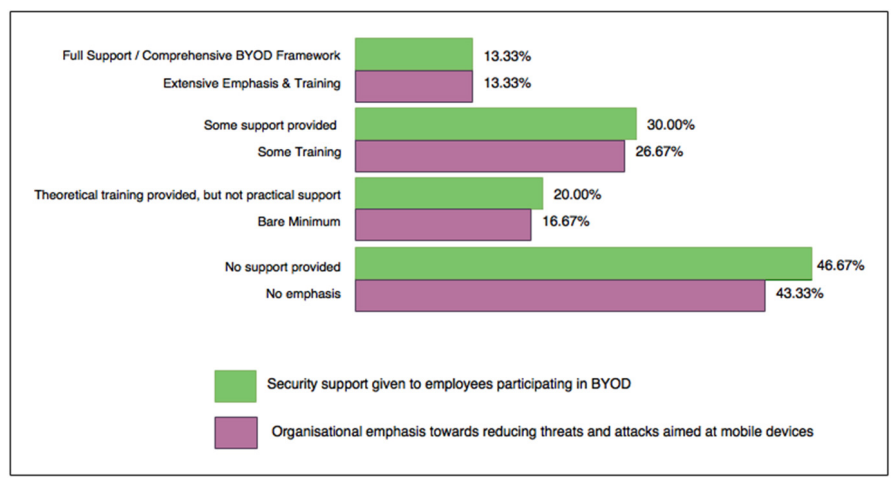

**Figure 2.** Organisational support for BYOD security compared to emphasis towards reducing attacks and threats in IT networks.

Employees' job roles do not necessarily influence the security controls applied or the resources and data accessed from individual devices, as 26.67% of studied companies consider them and 50% do not. The remaining survey respondents were unsure of this aspect, probably because employees are unaware of the network security controls applied, despite role-based access control being one of the most commonly applied security features in Australian business networks. These answers contradict commercial frameworks and academic recommendations for role-based data and resource access on a per user and device basis. This suggests that Australian businesses are using alternatives outside of commercially available BYOD security frameworks such as Bradford's 10 Step Model and Oracle's Mobile Security Suite [41,56], which focus on role-based access. According to survey results, not enough Australian business are focusing on the benefits of role-based access in BYOD security, such as the ability to track who is accessing resources and when in order to find out who is accountable when a security event occurs, or for vulnerability assessment and metrics gathering to determine which resources need updating. By comparing the number of employees accessing data from cloud resources to other online information distribution services (Figure 3), it is obvious that Australian businesses are becoming more reliant on and comfortable with sharing information online due to convenience, simplicity, and low maintenance. This is indicative of the need to strengthen security controls to combat threats such as data leakage, internal sabotage, and the exploitation of company data.

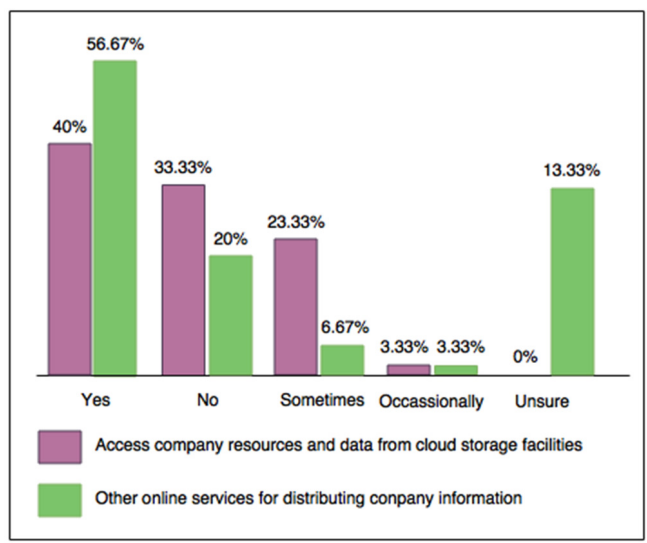

**Figure 3.** Data access from cloud storage and other online sharing services.

It is worth noting that BYOD security can only be effective if an organisation is protected by a robust network security policy to start with. For example, poor border protection or access control policy may render BYOD security ineffective. According to our study, types of network policies that are enforced for BYOD security, according to end users, presented predictable results, as pictured in Figure 4. A high rate of authorisation and authentication policies is expected, especially since most mobile devices enforce such methods by default. Today's BYOD-specific network design strategy may also need to take into account whether devices such as home routers and Internet of Things devices should be considered under the BYOD umbrella. Remote workers must set up secure router connections; otherwise, virtual private networks (VPNs) should be considered. In some cases, a guest network may be set up for unsecured BYOD devices to segregate unsafe personal devices from the enterprise network. Businesses could benefit if they also enforced these measures strictly as part of their BYOD security frameworks. Reporting and compliance policies are not common yet are just as important as the other types of policies, which exposes a flaw to address. Reporting policies may organically grow as a result of BYOD framework implementation growth. In Figure 4, 'Other' refers to organisations that use local storage encryption and inactivity time-outs (automatic session termination after a specific amount of time).

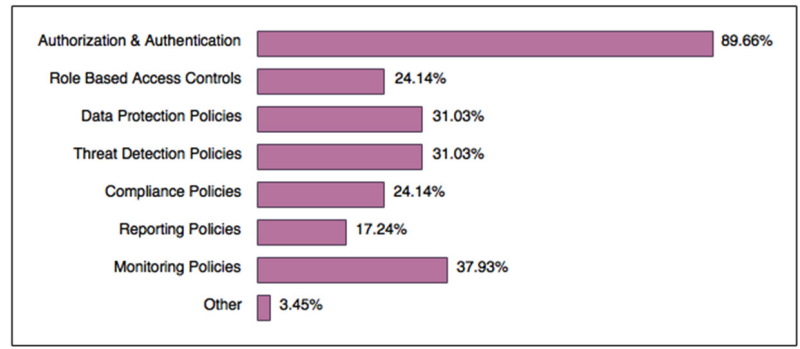

**Figure 4.** Network policies enforced by Businesses for BYOD security.

*4.3. User Practices in Regard to BYOD at Work*

In order to improve information security in business, it is vital to consider how end users actually engage with their devices to carry out work and how their activities relate to and determine the success of the security methods applied.

Knowledge of the applications employees commonly use to complete work tasks provides a scope with which to focus on security. Figure 5 granularly represents which applications employees use regularly to complete work tasks from BYOD-enabled devices. Applications concerning communication are most commonly used and this is due to people primarily relying on their smartphones, especially when working remotely, as these devices are specifically designed to conveniently facilitate real time communication. BYOD-enabled devices worldwide are most commonly used for accessing and storing work-related emails, scheduling information, documentation, web browsing, and social media [51,55], which is a global norm. Hence, there is a need for BYOD security frameworks to include policies that handle the stable and secure transmission of data when using these applications, such as encryption techniques. Compliance and data protection policies should also be active to protect data and files stored in applications and cloud servers.

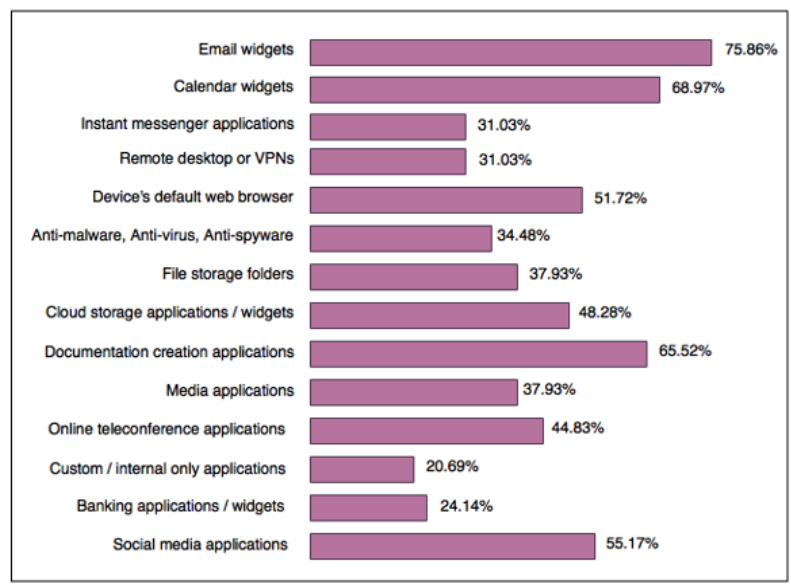

**Figure 5.** Applications regularly accessed via mobile devices for work purposes.

In total, 86.21% of the surveyed employees use applications installed on their mobile devices for both work and personal reasons. Another 53.85% admitted to sometimes using company resources for personal reasons because BYOD made it convenient, whereas 19.23% have only done this a few times. Furthermore, 65.52% of employees permit web browsers to save login credentials for websites on their devices (38.48% claimed they do not). These statistics should be perceived as a warning that internet security needs to be at the forefront of any BYOD security framework. Prescribed security mechanisms, which can reduce the probability of risks associated with these issues, such as data leakage, security breaches, and vulnerabilities, and create a clear divide between personal and private use, include containerisation and desktop virtualisation models [57]. Most surveyed employees care enough to utilise the security controls provided by modern computing devices on their own accord; mainly those enforced by default, such as PIN /Pass codes, passwords, or lock screen patterns (72.41%, 31.03%, and 20.69% of users respectively). Compared to a study conducted in Asia, which claimed that a majority of users secure devices with a combination of PIN codes and passwords and automatic screen locking, Australians are equally like-minded. Despite this good news, 10.34% of survey respondents are still not using any security methods at all, which can be reduced if businesses and educational institutions emphasise the importance of basic security features on mobile devices. It is possible this included people with older devices that do not enforce security measures by default. Relatively speaking, when asked if end users know and understand the potential risks and threats that affect mobile devices, 37.93% claimed they were well educated about the topic and 34.48% expressed that they had some knowledge. Some of the well-educated respondents are those whose jobs are centered around computing technology, such as software engineers. On the contrary, 17.24% of employees admitted they have a limited knowledge of problems faced by mobile devices, whilst another 10.34% had no knowledge of threats and attacks aimed towards mobile devices. The similarity amongst them was that these employees also answered to a lack of formal BYOD security agreements and training for device security. It is unclear in this survey which threats they need more education on, therefore it helps if businesses conduct vulnerability analysis to find out exactly where knowledge is poor and provide training accordingly. A factor that may contribute to the chance of security breaches occurring is that 79.31% of survey respondents admitted their company does not limit the use of BYOD outside of contracted work hours, though further exploration of this relationship is required.

Permissions for storing data on employee devices and the associated conditions varied between the organisations referenced in this study. Conditions were categorised as:

- It is permitted, though it is heavily monitored by MDM, MIM, or other strict monitoring policies—10.34%;
- The company trusts employees enough to allow data to be stored on devices without supervision—31.03%;
- Only publicly accessible data and resources are permitted on mobile devices—10.34%;
- This condition is forbidden for all staff—20.69%;
- No permission is granted, yet employees sometimes do this anyway to finish tasks after hours—6.9%.

Considering that a majority of employees store business-related data on their devices, firm data protection and threat detection policies need to be a priority for BYOD security frameworks in order to prevent data loss or the transmission of corrupted files into a company's internal network.

An estimated 68.97% of respondents engaging in BYOD claim that they or another co-worker has not experienced any security breaches whilst engaging in BYOD initiatives. If this is accurate, it means a decent percentage of organisations have avoided security breaches thanks to the introduction of BYOD security mechanisms, proving that it is more important than ever to consider these options. Another 20.69% claimed they were unsure if anyone else in their company had experienced a security breach on their mobile device. However, 10.31% admitted they or a co-worker have experienced a security incident with their device, and the events that occurred were primarily viruses. Users generally reacted the same way, by downloading antimalware software on their device to eliminate the issue. Survey respondents who experienced a security breach also belonged to companies with no to very little training, complete reliance on authorisation and network access control policies, and an absence of formally signed usage agreements.

*4.4. User Perceptions of BYOD in the Workplace*

When questioned on whether the introduction of BYOD security practices had improved working conditions, 48.15% of respondents admitted it had, to varying degrees, meaning employees appreciate the advantages of BYOD security frameworks. Some positive effects include:

- More confidence interacting with sensitive data and resources—42.86%;
- More trust is placed towards the organisation's processes—42.86%;
- Some resources and data are easier to access as a result—19.05%;
- Some work processes are more efficient since BYOD security measures were implemented—23.81%;
- Relief that staff is monitored and controlled more closely—14.29%.

A growing reliance on BYOD capabilities due to the convenience of mobile phone applications means that more people are working overtime, as 25% claimed this was a direct consequence of BYOD, whilst 52.17% said sometimes or occasionally, and 17.86% said they have never worked overtime due to BYOD initiatives. Work–life balance for the average Australian employee is becoming blurred, and this consequence could be an ethical concern for future BYOD frameworks. When survey participants were asked if they feel their privacy is invaded as a consequence of active BYOD security measures, 14.81% said 'yes', 48.15% said 'not at all', and 37.04% said 'sometimes', though this depends on the conditions and resources they are accessing. Generally, employees accept monitoring and other security policies providing that they only affect work applications. Reasons why people feel a sense of privacy invasion may be because they do not know the extent of the monitoring policies, paranoia or feel violated, feel that their company does not trust them, or because of the negative reviews of BYOD security mechanisms such as MDM, which are perceived as invasive amongst some users [44]. Some of the negative implications of BYOD security, which correspond with surveyed employees' concerns for personal privacy include:

- BYOD security policies have restricted access to certain resources and data too much—28.57%;
- Some methods have forbidden staff to use resources and data they once had access to—14.29%;
- Security methods used for certain resources are too excessive—14.29%;
- Employees tend to avoid some resources and data because of the security measures applied to them—35.72;
- Security methods make access to resources too inconvenient—35.71%.

Australian employees, according to this study, agree with security practices and policies enforced by their company network (74.07%), signifying that companies are already incorporating security measures that are well balanced and are fair for all affected stakeholders, though respondents disagreed with some policies for these reasons:

- Some policies or procedures are too restrictive; therefore, it is inconvenient to complete certain tasks—7.41%;

- They are too biased towards organisational interests and disregard employees—7.41%;
- A policy does not suit my department or job role requirements—3.70%;
- A policy is difficult to understand; therefore, there is resistance—3.70%;
- The remaining 18.52% of respondents expressed the need for more BYOD-specific policies, especially in regard to data access and to ensure that security methods are compatible with their mobile devices.

This study revealed some reasons why Australian employees dismiss certain security policies enforced for BYOD security, which provides further insight than previous studies which just claimed that, on average, 36% of employees outright ignore their company's policies [6]. In relation to this, 3.7% of surveyed employees admitted they do not adhere to some policies and practices due to the reasons listed, whilst 11.11% sometimes ignore certain policies because they were a nuisance. In contrast, 74.07% obey all policies and practices regardless. Although only a minority of employees disregard some policies, companies are advised to review them routinely to determine how necessary they are and whether to amend, overlook, or remove them. Business analytics tools can be used to measure the effectiveness of security policies, as they quantifiably present results to help analysts objectively decide how to proceed. The most insight was gained when people were asked if they have any suggestions for improving BYOD conditions and security in the workplace. Figure 6 displays the most useful responses.

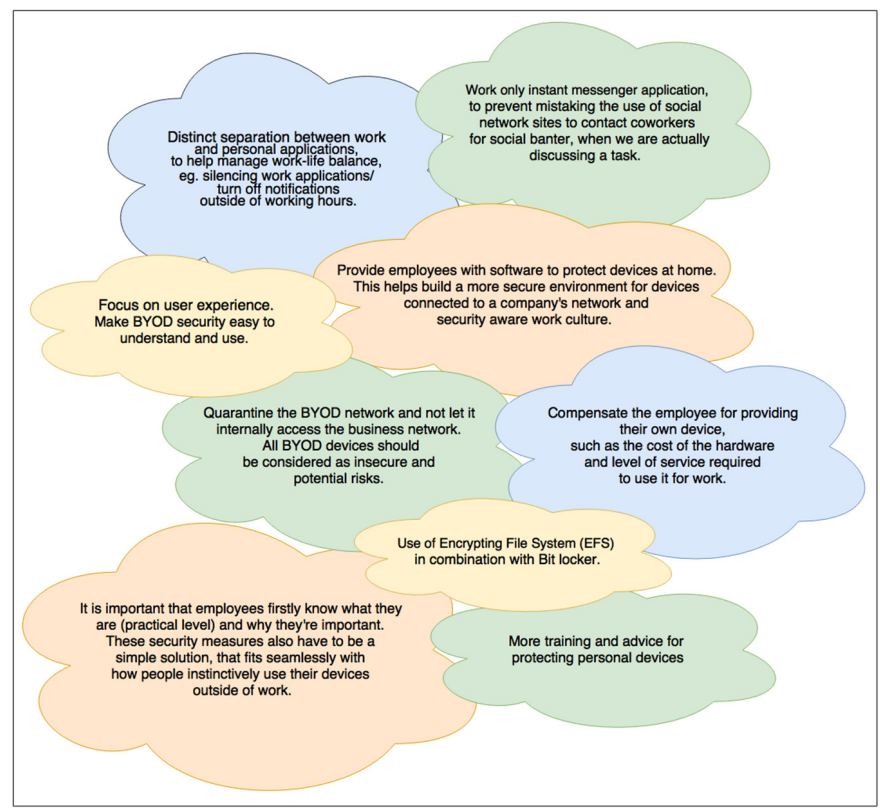

**Figure 6.** Employee suggestions for improving BYOD security at work.

## 5. Analysis and Discussion

This section analyses the responses and highlights patterns found in the survey results to confirm the reality of BYOD security conditions in Australian businesses and then discusses how these findings answer the three main research questions.

*5.1. Awareness of BYOD Security Aspects*

This study demonstrates that Australian businesses are tantamount to studies based on international demographics according to survey respondents in regard to BYOD-specific security mechanisms. Remote wiping has a consistently low usage rate in Asia, according to a survey conducted in India [55]. This study also confirmed that Australian businesses (approximately 20%) are equally as uninterested or unaware of remote wiping. The lower popularity levels of new mechanisms such as MDM, MAM, and MIM, remote wiping, and containerisation in Australia might be due to deficient research before implementing formal BYOD policies or the avoidance of overstepping privacy boundaries or investing time and money into implementing these frameworks. However, published literature has stated that people are partaking in BYOD without company knowledge, or prematurely for workplace preparedness [6,55]. As a result, companies depend on employees to protect their own devices [5,22,53]. Therefore, it is more likely that Australian employees have started using devices prematurely, thus forcing companies to strengthen existing security methods in their IT networks on an impromptu basis. Positively, compared to a study conducted by Cisco in America reporting that 40% of users do not use passwords on their devices [57], Australians are aware of how effective authorisation techniques are for data protection, as this survey recorded 89.66% of employees relying on one or more of the default methods provided by devices. A study conducted in Asia claimed that antimalware is also suffering from low popularity rates [55], yet survey respondents from this study indicated the opposite for the Australian workforce (almost 70%), who favoured this security mechanism most over other options.

This study detected a relationship between employees' job roles and levels of reliance on BYOD, as depicted in Table 5. Job roles classified as management positions tend to be more reliant on BYOD capabilities than other roles. Upper-management roles such as CEOs and CIOs are typically the most concerned with BYOD security and its side effects on business operations and revenue [9]. Managers are welcoming the opportunity to utilise technology to supervise and engage with staff more closely, and as such more responsibility is placed on them to set a good example for how BYOD strategies can be utilised safely. For IT specialists, BYOD acceptance is a no-brainer, especially for system and network administrators who rely on it to conduct maintenance and troubleshoot problems offsite or after hours. In the case of this survey, 'other' job classifications consisted of practical trades, such as healthcare, engineering, etc., which tended to be less reliant on mobile devices for everyday work operations.

**Table 5.** Job role classification relationship with reliance on BYOD strategies.

| Job Classification | Extremely Reliant | Highly Reliant | Moderate Use | Casual Use |
|---|---|---|---|---|
| Management | 14.28% | 11.43% | 11.43% | 0% |
| Information Technology | 5.72% | 8.58% | 8.58% | 2.86% |
| Other | 2.86% | 8.58% | 14.28% | 8.58% |

Most survey respondents (65.52%) allow web-based applications to save login credentials, which was expected considering that 88.57% use their privately owned devices for work at varying degrees and over 85% use the same applications for both work and personal reasons. People prioritise convenience and easy ways to communicate, yet if a device is lost or stolen, there is a chance that confidential business data can be exploited or used to damage the company's reputation [58,59]. Businesses can increase their staff's awareness of security using simple training techniques such as quarterly reviews about security topics, which are conducted like company-wide training sessions, or as short presentations.

*5.2. Employee Responses to Security Mechanisms Applied for BYOD*

Considering that 14.81% of employees expressed a strong concern for the invasion of their personal privacy, a compromise between this and organisation's needs cannot be overlooked. These feelings of privacy invasion can be relieved with:

- A monitoring policy that pauses or turns off BYOD security agents installed on mobile devices when employees are not working.
- Explicitly publishing in usage and liability agreements exactly what is monitored, why and when, and provides descriptions of activities they may be investigated as suspicious [23].

Over 50% of employees believe that the introduction of BYOD security has improved their working conditions, did not feel as if their privacy is invaded, and would not mind if their organisation monitored only work applications installed on mobile devices. Most employees agree with the policies in place for BYOD security and adhere to them. This combination of factors encourages organisations to consider and be reassured that employees are willing accept BYOD security policy-based frameworks. However, to experience a successful deployment, it is recommended to enforce policies that only affect work related applications and that are less likely to be perceived as too invasive towards personal privacy. Half of the respondents who felt like their privacy was invaded (7.4%) have MDM agents installed on their mobile devices; which have the stigma of being highly invasive. In contrast, out of the total respondents that felt like their privacy was sometimes invaded, MDM was considered equally as invasive as other methods. This is good news for companies considering MDM, as Australian employees seemingly place more trust in their organisations compared to Europe and America, where publications stated that employees showed resistance towards MDM [44,60].

Majority of employees (73.08%) have used a company resource for personal use (downloads, casual conversations, etc.) occasionally or sometimes, due to the convenience of BYOD initiatives. Associated potential risks such as data leakage, internal espionage and the opportunity for malware invasions need to be accounted for from a security perspective. Companies need to be aware of how monitoring and data protection techniques meet requirements of local privacy laws. For example, internet monitoring may be limited to recording how many times and how long a session lasted for each website visited by an employee during work hours.

A portion of survey respondents expressed concerns regarding hardware specifications required for facilitating security mechanisms on their devices. Ideally, BYOD frameworks should cater to hardware with limited storage space and be efficient in terms of processing power. If side effects, such as, poor latency, instability, overheating or fast depletion of battery life occur, employees will be resistant. A suggestion to relieve these is by placing a time limit for when the framework's mobile agent is active within the device; e.g., It can only be accessed between certain hours of the day and use containerisation to ensure applications are only active when the user has created an active session inside the container.

*5.3. Weaknesses Affecting BYOD Security From the End Users Perspective*

Survey results regarding the use of formally signed BYOD policies highlights a critical need for BYOD security frameworks to incorporate and continually reinforce policies, such as, signed acceptable usage, user agreement policies and liability contracts. Secure Mobile Business Framework (SMBF) heavily enforces this as its first line of defence, by ensuring employees sign and agree to practically applied network security policies that will be active for all BYOD enabled devices [36]. Existing and future BYOD security frameworks are advised to borrow this concept from SMBF in order to ensure this weakness diminishes, as well as strengthen countermeasures against the ever-growing threats targeting mobile devices. Survey respondents demonstrated a strong desire to have prominent formal BYOD security policies, as a number of them specifically suggested this need when asked for their opinions for improving BYOD security at work. Australians agree with employees spread globally, where approximately 97% of organisations believe that mobile device security policies are very important [23,53]. It has been stated that companies who have formal BYOD policies are not refining them enough, are not clearly communicating them to employees or are ignored [6,51,57], which might explain why 13.33% of survey respondents were 'unsure' when asked if their workplace had formal BYOD policies in place. Some recommended remedies for enhancing security policies include:

- Explaining simply in acceptable usage and user agreements the goals of BYOD specific security policies, expected behaviours, permitted activities, and prohibited access rules whilst participating in BYOD initiatives [5].
- Incorporating the ISO 27001–27006 IT security standards, which provide guidance for professional security practices, a general code of practice for portable devices and managing human resources [6,51,61].
- Base the aims of attack countermeasure policies on risk and threat assessments findings [51].
- Consult with employees when deciding policies for departments, in order to ensure fairness and that they will be educated and willingly follow security process simultaneously [10,61].

Lack of training and support provided by Australian businesses requires heavier emphasis in regard to BYOD security and provide support to protect all mobile devices present in the workplace. Companies cannot assume employees will independently protect their mobile devices and the networks they connect to, if there is an absence of training that specifies even basic security mechanisms and threats to be aware of. Survey results indicate that majority of employees rely on the authority of their governing business to protect their mobile devices whilst at work. Additionally, a portion of survey respondents expressed a need for more training and advice for independently securing personal devices to help decrease the chances of security threats for all company stakeholders, as demonstrated in Figure 6, Section 4.

The high reliance on social media, cloud based, and other services hosted online for the purpose of information sharing also highlights importance of formally signed policies. It is imperative that data protection and monitoring policies are in place to manage security when accessing cloud based and online based communication facilities. For example, internet monitoring can deter authorised users from using information unethically as well as gain access to potentially unsafe websites. Online information sharing, when combined with BYOD multiplies risks such as data leakage threefold [55,58]. When using third party software or cloud-based storage, companies compromise control of where and how data is stored and accessed by these applications, thus more responsibility is placed on the business to independently apply security controls. Such storage solutions generally rely on basic authentication methods such as username and password credentials, which can be bypassed by hackers [62]. Policies and security methods such as MIM and MAM, which enhance protection of applications and information transmitted through them, can help companies regain extra control over security of externally stored data.

Results gathered regarding the network security policies enforced compared to authorisation granted enabling employees to store data on devices, shows that not enough is being done to protect devices, as only 10.34% of businesses enforce strict data protection, yet approximately 59% of employees are storing work related data on their devices with or without company permission. This is becoming a prevalent weakness for existing BYOD security frameworks worldwide [57]. Security risks are higher for mobile devices containing company data without data protection policies and mechanisms such as remote wiping and antimalware [55], especially if they are lost or stolen, which is a primary concern for BYOD security [62]. In the scenario that a device is stolen, the data can be exploited and leaked due to effortless unauthorised access [55], resulting in the loss of intellectual property for the original owner.

**6. Limitations of the Current Study**

This study was centered around the success of an online survey, which presents some implications.

Specifically targeting an Australian audience created a relatively smaller niche of respondents who were able to participate. However, given that Australian businesses are still investigating or are just starting to introduce formal BYOD strategies, those who responded to the survey are likely to be the most legitimate sources of this research data. Human interpretation of survey questions affected how survey respondents understood what questions were asked of them, and may have been misinterpreted, or an unwillingness to share certain details could have affected the overall accuracy of survey results. In relation to questions, the questions that were deemed potentially sensitive for respondents, offered answer options that allowed respondents to opt out of giving legitimate answers if they felt uncomfortable. Whilst these factors were considered during survey analysis, there was no way to determine the exact extent of the accuracy of the data returned.

## 7. Conclusions

This research verified key points previously discovered in existing literature, as well as illuminate new considerations and inspired ideas for improving existing and future BYOD security frameworks in regard to Australian Businesses, from the perspective of surveyed employees. The conducted survey revealed some surprising facts in Australia's approach to BYOD security; firstly, approximately 40% of Australian businesses referenced enforce formal BYOD specific security policies, which is more or less on par with international standards. Preferred security methods are those that have been prevalent in IT networks since before the introduction of smart phones (most popular mobile devices for BYOD), such as network access controls and antimalware. Employees particularly rely on applications for communicating with co-workers and clients, documentation and planning schedules. Managers and information technologists are leading the trend as the most reliant on BYOD, whilst the retail and telecommunications industries have also been the most accommodating of BYOD strategies in the workplace. The assimilation rate of BYOD security mechanisms is still developing in Australian businesses, although awareness of newer BYOD security frameworks still requires growth. With more publicity about this topic, the rate of businesses implementing more substantial BYOD security frameworks could rise quickly.

Business and employee awareness of BYOD security aspects presented positive results. Nearly 90% of employees surveyed use at least one of their device's default security mechanisms, and a majority of employees believed that they had an adequate to good knowledge of the potential threats and risks targeting mobile devices. Employees' practices and perceptions towards the security mechanisms applied for BYOD are reassuring to other businesses that are still debating the importance of BYOD in modern work environments. Overall, employees welcome and desire stable BYOD security frameworks for the sake of protecting their personal mobile devices, as most agree that BYOD security has introduced positive changes. Weakness in BYOD security includes blurred boundaries between work and personal applications and lack of signed and written policies and agreements. Enforcing these policies will enhance employees' knowledge of cyber risks and encourage them to help the business by independently protecting mobile devices, which can reduce data leakage and contamination and the spread of malware. It is recommended for businesses to examine employee practices and perceptions and consult departments in order to maintain BYOD security frameworks that are most efficient for unique security requirements. It is advisable for modern businesses to rely more on multi-purpose BYOD-specific mechanisms such as MDM, MIM, and MAM, as they include specialised reporting, data protection, and monitoring policies for mobile devices, as well as in-built remote wiping functions. Future considerations need to focus more on combating the side effects of lost, stolen, and obsolete devices and on the storage of confidential data on hardware memory and internet-based services.

**Author Contributions:** Writing—original draft, K.D.; Writing—review and editing, M.B. All authors have read and agreed to the published version of the manuscript.

**Funding:** This research received no external funding.

**Institutional Review Board Statement:** The study was conducted in accordance with the Declaration of Helsinki, and approved by the Human Research Ethics Committee of Charles Sturt University, Australia.

**Informed Consent Statement:** Implied informed consent was obtained from all subjects involved in the study.

**Data Availability Statement:** Not available in the public domain.

**Conflicts of Interest:** The authors declare no conflict of interest.